\documentclass{article}
\usepackage[utf8]{inputenc}
\usepackage{authblk}
\usepackage{setspace}
\usepackage[margin=1.25in]{geometry}
\usepackage{graphicx}
\graphicspath{ {./figures/} }
\usepackage{subcaption}
\usepackage{amsmath}
\usepackage{lineno}
\usepackage{amsmath}
\usepackage[colorlinks=true, allcolors=blue]{hyperref}
\usepackage{bm}
\usepackage{float}
\usepackage{color}

\usepackage[style=nejm, 
citestyle=numeric-comp,
sorting=none]{biblatex}
\title{\fontsize{16}{14}\selectfont\textbf{Slimmed Optical Neural Networks with Multiplexed Neuron Sets and a Corresponding Backpropagation Training Algorithm}}

\author[1,2]{Yi-Feng Liu}
\author[1]{Rui-Yao Ren}
\author[1,2]{Dai-Bao Hou}
\author[7]{Hai-Zhong Weng}
\author[8]{Bo-Wen Wang}
\author[1]{Ke-Jie Huang}
\author[1,2,3]{Xing Lin}
\author[1,2,3,4]{Feng Liu}
\author[1,6]{Chen-Hui Li}
\author[1,2,3,4,5]{Chao-Yuan Jin \thanks{Corresponding author: jincy@zju.edu.cn}}

\affil[1]{College of Information Science and Electronic Engineering, Zhejiang University, Hangzhou 310027, China}
\affil[2]{State Key Laboratory of Extreme Photonics and Instrumentation, Zhejiang University, Hangzhou 310027, China}
\affil[3]{Interdisciplinary Center for Quantum Information, Zhejiang University, Hangzhou 310027, China}
\affil[4]{International Joint Innovation Center, Zhejiang University, Haining 314400, China}
\affil[5]{Center for Information Technology Application Innovation, Shaoxing Institute, Zhejiang University, Shaoxing 312000, China}
\affil[6]{Zhejiang Lab, Hangzhou 311121, China}
\affil[7]{School of Physics, CRANN and AMBER, Trinity College Dublin, Dublin 2, Ireland}
\affil[8]{Synopsys, Inc., 7521 PL Enschede, the Netherlands}
\date{} 

\begin{document}

\maketitle
\renewcommand{\abstractname}{\large Abstract\\}
\begin{abstract}
{\fontsize{10.5}{14}\selectfont Optical neural networks (ONNs) have recently attracted extensive interest as potential alternatives to electronic artificial neural networks, owing to their intrinsic capabilities in parallel signal processing with reduced power consumption and low latency. Preliminary confirmation of parallelism in optical computing has been widely performed by applying wavelength division multiplexing (WDM) to the linear transformation of neural networks. However, interchannel crosstalk has obstructed WDM technologies to be deployed in nonlinear activation on ONNs. Here, we propose a universal WDM structure called multiplexed neuron sets (MNS), which applies WDM technologies to optical neurons and enables ONNs to be further compressed. A corresponding back-propagation training algorithm was proposed to alleviate or even annul the influence of interchannel crosstalk in MNS-based WDM-ONNs. For simplicity, semiconductor optical amplifiers are employed as an example of MNS to construct a WDM-ONN trained using the new algorithm. The results show that the combination of MNS and the corresponding BP training algorithm clearly downsizes the system and improves the energy efficiency by magnitudes of ten while providing similar performance to traditional ONNs.}
\end{abstract}


\section{Introduction}
Machine learning (ML) technologies have developed rapidly in recent years. Empirical evidence has shown that the capabilities of ML match or even exceed human intelligence in fields such as speech recognition, image classification, and intelligence-competitive games \cite{mnih2015human,silver2017mastering,butler2018machine}. With the ML technological boom , especially in artificial neural networks (ANNs), optical neural networks (ONNs) have become a potential part of the future infrastructure for ML and are believed to be a competitive alternative to their traditional electronic counterparts\cite{de2019photonic,xu2021survey,shastri2021photonics,huang2022prospects,zhou2022photonic,shen2017deep}. Because optical systems feature inherent parallelism with low energy consumption and low latency, the merging of electronics and optics is expected to alleviate some of the drawbacks of fully electronic systems\cite{liu2021research,sui2020review}. Regarding the two fundamental elements of ANNs, vector-matrix multiplication and nonlinear activation function have been proved to  both benefit from space and time division multiplexing in ONNs \cite{farhat1985optical,gruber2000planar,lin2018all,shen2017deep,zhang2021optical,qian2020performing,feldmann2019all,wang2022optical,refadv}. In addition, wavelength-division multiplexing (WDM), enabled by encoding information onto various wavelengths, provides an exclusive dimension of parallelism for ONNs. Therefore, preferable performance have been achieved with off-the-shelf optoelectronic WDM devices \cite{ishihara2019optical,totovic2022programmable,mourgias2020all,shi2019deep,feldmann2021parallel,xu202111,huang2021silicon,tait2016microring,refadv}. 

Although remarkable efforts have been made at both the hardware and software levels for a slimmed ONN, the focus of WDM technologies applied to ONN has been limited to the vector-matrix multiplication part \cite{zhao2019hardware,wang2022metasurface,zhu2022space}. As for optical-based nonlinear activation functions, various optoelectronic devices, such as semiconductor optical amplifiers (SOAs), ring resonators, and optical phase modulators, among others, have been proposed and experimentally investigated \cite{feldmann2019all,tait2019silicon,williamson2019reprogrammable,amin2019ito,shi2022inp}. However, the nonlinear response of these devices inevitably causes crosstalk between channels when WDM signals are applied. There is no universal plan for slimmed ONNs that involves multiplexing nonlinear neurons without downgrading its performance.

In this study, we propose a structure called \emph{multiplexed neuron sets} (MNS) and a corresponding back-propagation (BP) training algorithm. The combination of these two  can compress \emph{n} parallelly-deployed neurons into 1 with the help of WDM while maintaining the original performance. We take SOAs as typical examples for the implementation of the MNS. The corresponding BP algorithm was designed to overcome the performance degradation caused by crosstalk between wavelength channels in SOAs. A slimmed ONN constructed using an MNS was proposed and trained using the corresponding BP algorithm. The results demonstrated that the eliminated scale greatly improved the energy efficiency of the entire system. Although SOAs have been employed as a possible implementation of the MNS for simplicity, other photonic devices are potential elements for MNS if they satisfy the features described in the following sections. The designed BP algorithm is universally suitable for various ONN architectures with interchannel crosstalk.

\section{Materials and Methods}
\subsection{MNS structure and SOA-Based MNS} 
\label{sec:MNS-intro}

\begin{figure}[htb]
\centering\includegraphics[width=8cm]{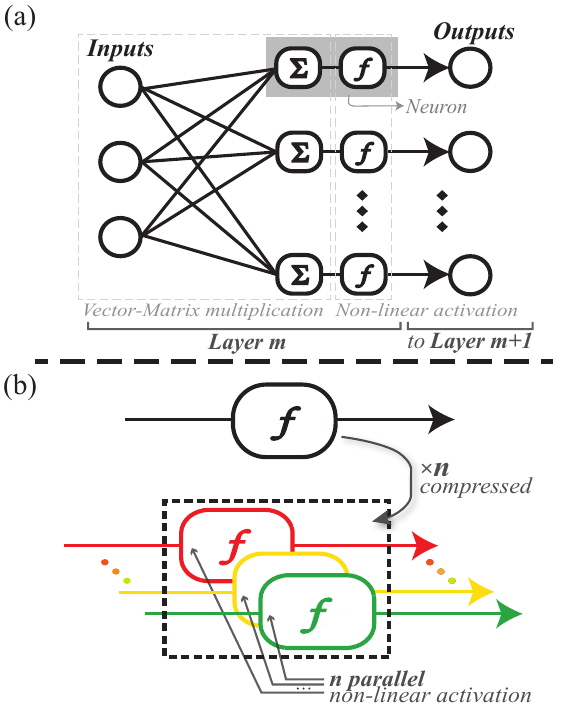}
\caption{(A) A scheme of a traditional FCNN; the layers are connected by the black lines, which corresponds to the weight matrix. The neurons separately realize the summation and nonlinear activation functions without influencing others. (B) An example of a nonlinear activation function and how it can be conceptually multiplexed in a single device.}
\label{FCNN}
\end{figure}

An simplified scheme of fully connected neural networks (FCNNs) is shown in Fig. \ref{FCNN}(A). The neuron marked in the gray-shadowed box acts as one of the basic elements of FCNNs. The propagation of data is realized through the full connections of the neurons in the adjacent layers. These connections, called synapses, have different weights and can be abstracted into a weight matrix that executes linear vector-matrix multiplications while the data are forward propagating. Neurons, however, execute summation ($\Sigma$) and nonlinear activation (\textit{f}) when they receive data from the previous layer. The summation function represents the last step of the vector-matrix multiplication, which is a part of the linear transformation. 

In conventional ONNs that deploy FCNNs, only one channel exists in each physical connection, which strictly represents one synapse, whereas the weights introduced by all synapses define the weight matrix. When WDM-ONN is applied, multiple wavelength channels (i.e., multiple synapses) are compressed into one physical connection. In the mathematical picture, each column of the weight matrix can be coded onto different wavelengths and subsequently compressed into one physical connection that virtually represents multiple synapses \cite{mourgias2019all,shi2019deep,xu2020photonic}, or otherwise each row of the weight matrix can be compressed \cite{totovic2022programmable,ishihara2019optical,feldmann2021parallel}. However, to the best of our knowledge, all compression approaches for WDM-ONN have been applied only to the linear transformation part of either the input vector or the weight matrix. 

Therefore, it is natural to assume that WDM can be deployed in nonlinear activation functions. Based on the concept illustrated in Fig. \ref{FCNN}(B), parallel activation functions are coded onto different wavelengths and executed in a single device, which is labeled in a dashed box. If the summation function ($\Sigma$) is multiplexed with the nonlinear activation function (\textit{f}), multiple neurons in the column in Fig. \ref{FCNN}(A) can be further compressed into one single functional unit, which we name multiplexed neuron sets (MNS). The ultimate goal of MNS is to simplify the system by implementing several summation and activation functions using a single photonic or optoelectronic device. Thus, at the network level, a device is multiplexed to act as multiple neurons. 

In Fig. \ref{MNS}(A), \emph{Layer m} is decomposed by a weight matrix and MNS. The corresponding physical structure of \emph{Layer m} is emphasized in the gray box. The input of the MNS structure in \emph{Layer m} is a vector resulting from the vector-matrix multiplication in \emph{Layer m} and is encoded by the input power of the MNS channels with various wavelengths. In this study, we provide an example of MNS realized using an SOA as shown in Fig. \ref{MNS}(B). The reasons for choosing SOAs as examples are as follows. 

\begin{itemize}
    \item SOAs are commercially mature devices and have become easy to access;
    \item SOAs' intrinsic characteristic of gain saturation have been employed as nonlinear activation functions elsewhere \cite{shi2022inp,mourgias2019all};
    \item It is suitable for SOAs to process multiple inputs encoded on various wavelengths in parallel; 
\end{itemize}

\noindent A MUX is used to combine different wavelengths and send to an SOA which will offer multichannel amplification. At the output port of the SOA, a DEMUX is used to split the outputs into separate channels. A set of nonlinear activation functions is applied between the inputs and outputs of each channel.

At the rightmost of Fig. \ref{MNS}(B), a list of multiwavelength channels  entering the SOA in parallel with various power levels are shown, whose power corresponds to the input of an individual neuron (separated by different colors) in Fig. \ref{MNS}(C). The power levels at the SOA output ports intrinsically represent the calculated results for the input at the same wavelength. For an ONN architecture containing a device that satisfies the features shown in Fig. \ref{MNS}(C), the concept of MNS naturally helps to scale down the number of devices in use. However, the nonlinear response of this device inevitably causes crosstalk between wavelength channels. Crosstalk can cause propagation errors, resulting in performance degradation. We believe that this obstructs the deployment of WDM in nonlinear activation functions in practice. For devices such as SOAs, crosstalk has been a disadvantage for their application in ONNs\cite{shi2022inp,totovic2022programmable}. Every input channel contributes to the gain-saturation effect, and the output signals suffer from amplification deviations. In other words, the output signal of each channel is determined not only by the input of this channel but also by the input of other channels. This phenomenon, which is induced by the gain saturation effect, is generally called cross-gain modulation. A compact model for  cross-gain modulation working at a relatively low modulation rate can be expressed as follows:

\begin{figure}[htbp]
\centering\includegraphics[width=8cm]{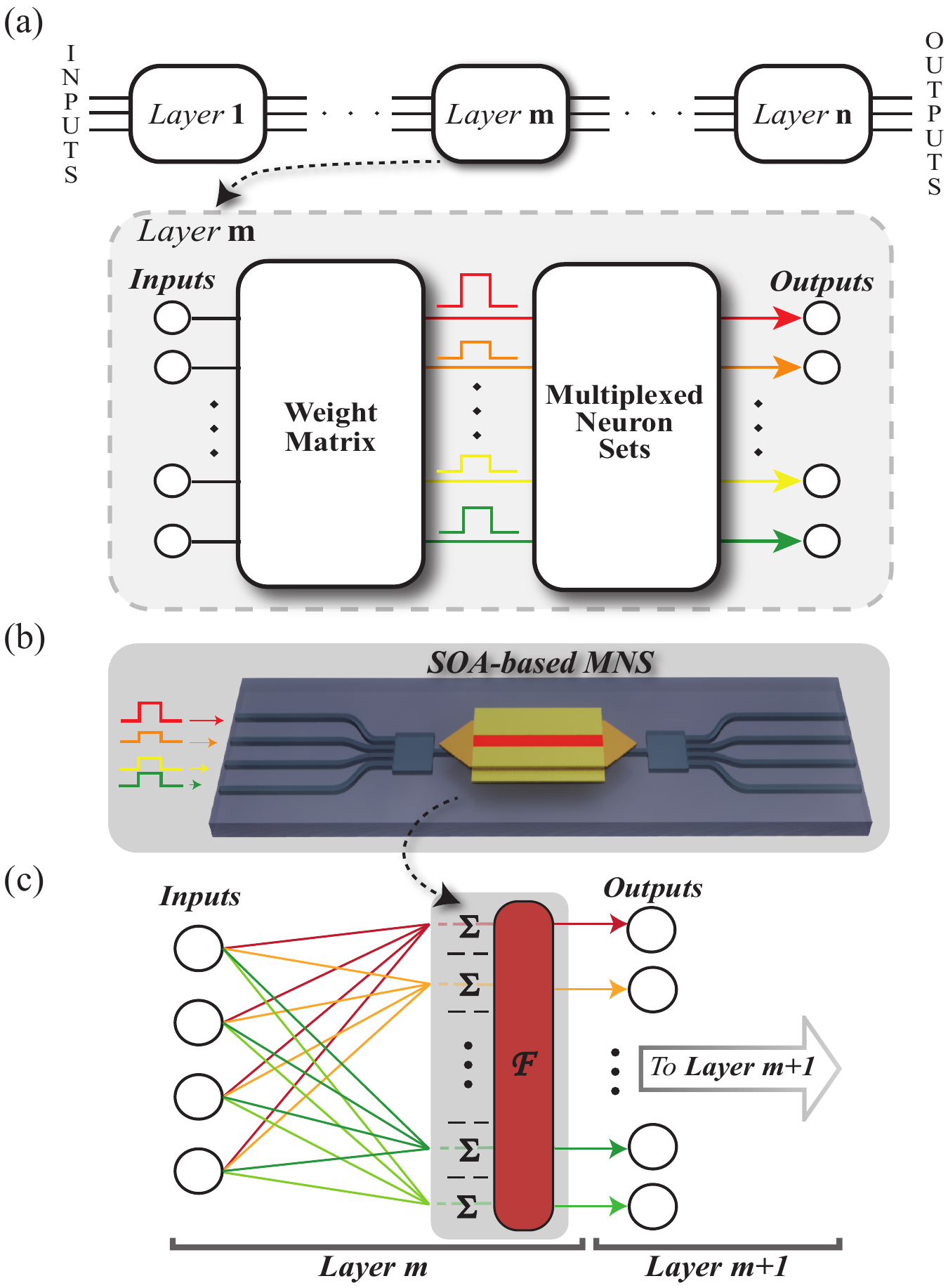}
\caption{(A) A block diagram of a WDM-ONN with an MNS structure. Multiple neurons are encoded on various wavelengths and input into MNS. (B) The MNS structure in this work is realized by a multichannel SOA. (C) A schemed connection picture for a WDM-ONN with a hidden layer composed of MNS.}
\label{MNS}
\end{figure}

\begin{equation}
G = \frac{{G_{ss}}}{{1 + \frac{{{P_{in}}}}{{{P_{sat}}}}}}
\label{gain-saturation}
\end{equation}

\noindent where $G$ and $G_{ss}$ are the single-pass gain and small-signal single-pass gain of the SOA, respectively, and $P_{sat}$ is the saturation power. As shown in Fig. \ref{MNS}(B), $P_{in}$ becomes the summation of a series of optical powers of various wavelength channels. The sum of the inputs can be expressed as

\begin{equation}
{P_{in}} = \sum\limits_{k = 1}^n {{P_{in\_k}}}
\label{input-sum}
\end{equation}

\noindent where $P_{in\_k}$ represents the input power of the $k^{th}$ channel. For simplicity, the wavelength dependence of the single-pass gain is ignored. 

When an excitation inputs the SOA, the steady state is reached fairly quickly, and the gain recovery time is usually on the timescale of nanoseconds. In other words, if we sample an SOA with a time duration much longer than a few nanoseconds, the nonlinear process inside the SOA will not cause severe frequency instabilities. As we have the input power of each channel and a single-pass gain, it is easy to calculate the output power of each channel.

\begin{equation}
{P_{out\_k}} = {P_{in\_k}} \times {G}.
\label{output-k}
\end{equation}

\begin{figure}[htb]
\centering\includegraphics[width=8cm]{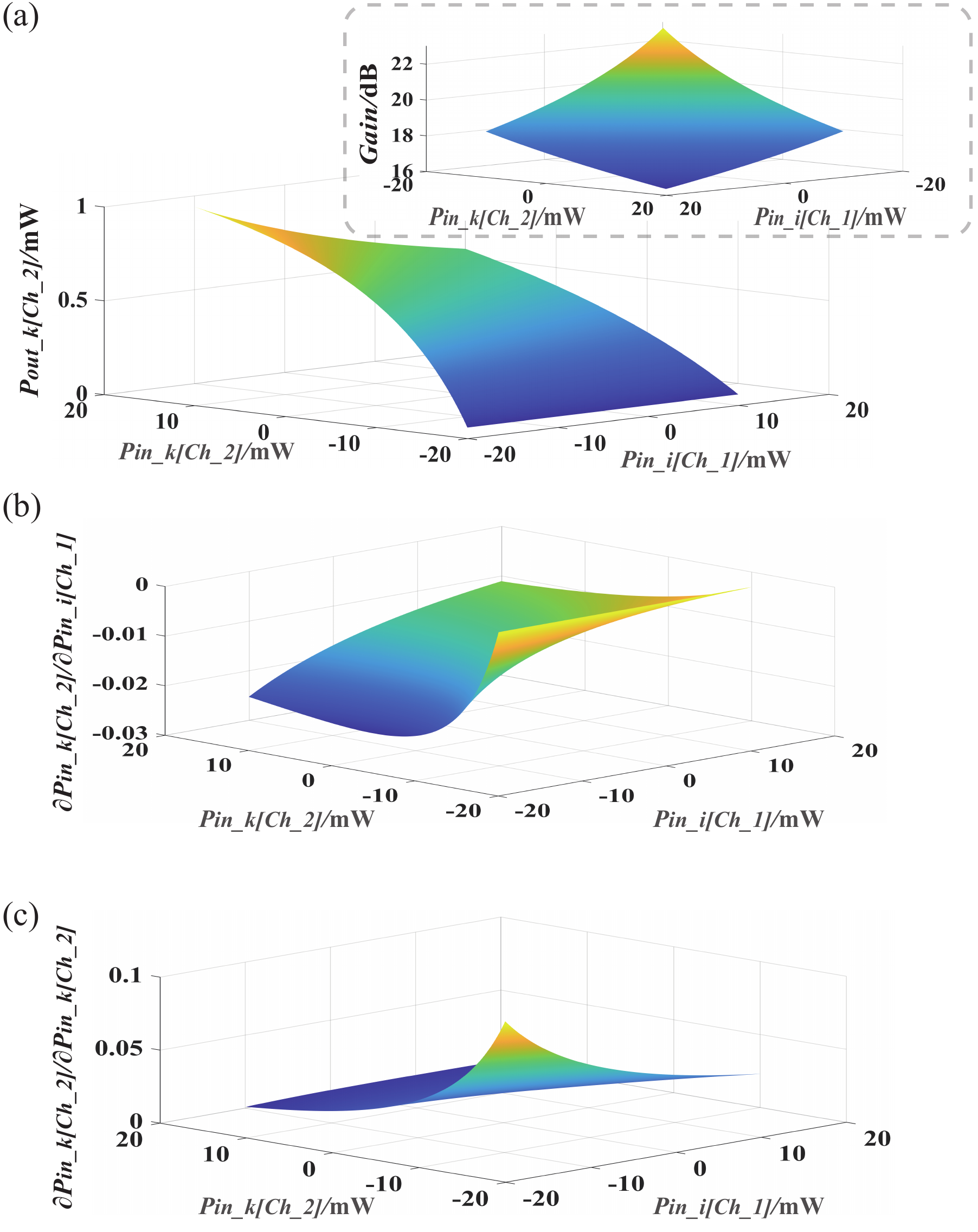}
\caption{For a 2-channel SOA. (A) Here, the output of \emph{Ch-2} versus the input of \emph{Ch-1} and input of \emph{Ch-2}is visualized. The inset shows the overall gain versus the input of \emph{Ch-1} and the input of \emph{Ch-2}. (B) The partial derivatives of the output to the inputs is visualized here: $\partial ({P_{out\_k}{[Ch\_2]}})/\partial ({P_{in\_k}}{[Ch\_1]})$. The partial derivatives of the output to the inputs is also visualized here:$\partial ({P_{out\_k}}{[Ch\_2]})/\partial ({P_{in\_k}}{[Ch\_2]})$.}
\label{2d-soa}
\end{figure}

For a more straightforward demonstration of the interchannel crosstalk, we provide an example of a 2-channel-multiplexed SOA in Fig. \ref{2d-soa}. The outputs of \emph{Ch-2} versus the inputs of \emph{Ch-1} and \emph{Ch-2} are shown in Fig. \ref{2d-soa}(A), and the overall variation in single-pass gain is shown in the inset. When the input of \emph{Ch-2} remains constant, the gain decreases as the input of \emph{Ch-1} increases; thus, the output of \emph{Ch-2} decreases. This is clear evidence of crosstalk between \emph{Ch-1} and \emph{Ch-2}. To investigate the influence of the inputs on the output further, we calculate the partial derivatives of the output. As shown in Figs. \ref{2d-soa}(B) and (C), $\partial ({P_{out\_k}{[Ch\_2]}})/\partial ({P_{in\_k}}{[Ch\_1]})$ and $\partial ({P_{out\_k}}{[Ch\_2]})/\partial ({P_{in\_k}}{[Ch\_2]})$ are plotted, as the partial derivatives are fundamentally important elements in BP training algorithm.

\subsection{The corresponding BP training algorithm}
\label{sec:bp-sec}

To enable the use of MNS in ONN, a new BP training algorithm was developed to alleviate or even annul the degradation caused by interchannel crosstalk. For an SOA with multichannel input, the output of each channel can be represented as a multivariable function, with the input of each channel as variables. The entire output vector, which is composed of the outputs of all the channels, is a set of multivariable functions that share the same input variables. For an n-channel SOA, the ${i^{th}}$ channel output can be written as

\begin{equation}
{y_i} = {x_i} \times ({\frac{{G_{ss}}}{{1 + ({x_1} + {x_2} + ... + {x_n})/{P_{sat}}}}})
\label{multi-SOA}
\end{equation}

\noindent here ${y_i}$ is the ${i^{th}}$ channel output and ${x_n}$ is the ${n^{th}}$ channel input. For simplicity and universality, we abstract the multivariable functions as ${y_i} = {f_i}({x_1},{x_2},...,{x_n})$, through which the MNS that are constructed by nonlinear optical or optoelectronic devices are unified in mathematical level. For both the output and hidden layers of the network, the output of a specific layer is a column of multivariable functions.  

During the training of a specific layer, the partial derivative of the loss $L$ with respect to the weight matrix $\bm{{W}}$ is calculated according to the chain rule. The corresponding new BP algorithm inherits the idea of minimizing the loss along the gradient direction while coupling the matrix below into the chain rule. 

\begin{equation}
{\frac{{\partial \bm{output}}}{{\partial \bm{{s}}}}}  = \left( {\begin{array}{*{20}{c}}
{\frac{{\partial outpu{t_1}}}{{\partial s_1}}}& \ldots &{\frac{{\partial outpu{t_n}}}{{\partial s_1}}}\\
 \vdots & \ddots & \vdots \\
{\frac{{\partial outpu{t_1}}}{{\partial s_n}}}& \cdots &{\frac{{\partial outpu{t_n}}}{{\partial s_n}}}
\end{array}} \right),
\label{ele-expand}
\end{equation}

\noindent here $\bm{{s}}$ represents the result vector of the vector-matrix multiplication of this layer. Note that this matrix represents the inner difference between the new BP algorithm and the traditional algorithm caused by crosstalk. It is evident that each element in the matrix has a definition corresponding to the crosstalk between the channels, as shown in Fig. \ref{2d-soa}. In a traditional BP algorithm, the elements on the diagonal have definitions, whereas the off-diagonal elements are left undefined. The new BP algorithm deals with physical crosstalk and couples mathematical operations to the undefined items of the traditional BP algorithm. [For a detailed derivation of the new BP algorithm, see Supplementary Information.]

Presented both in the previous section and in the Supplementary Information, the expressions for the crosstalk among the channels are highly parameterized and abstract enough to couple to various devices or even all types of crosstalk. Although the noncoherent situation for SOA-based MNS is explicitly presented in this work, the corresponding training algorithm still maintains the ability to handle the coherent situation as long as the outputs and inputs of the multiport nonlinear part of the network comply with the function ${y_i} = {f_i}({x_1},{x_2},...,{x_n})$.

\subsection{Crosstalk level evaluation in SOA-based MNS}
\begin{figure}[htbp]
\centering\includegraphics[width=8cm]{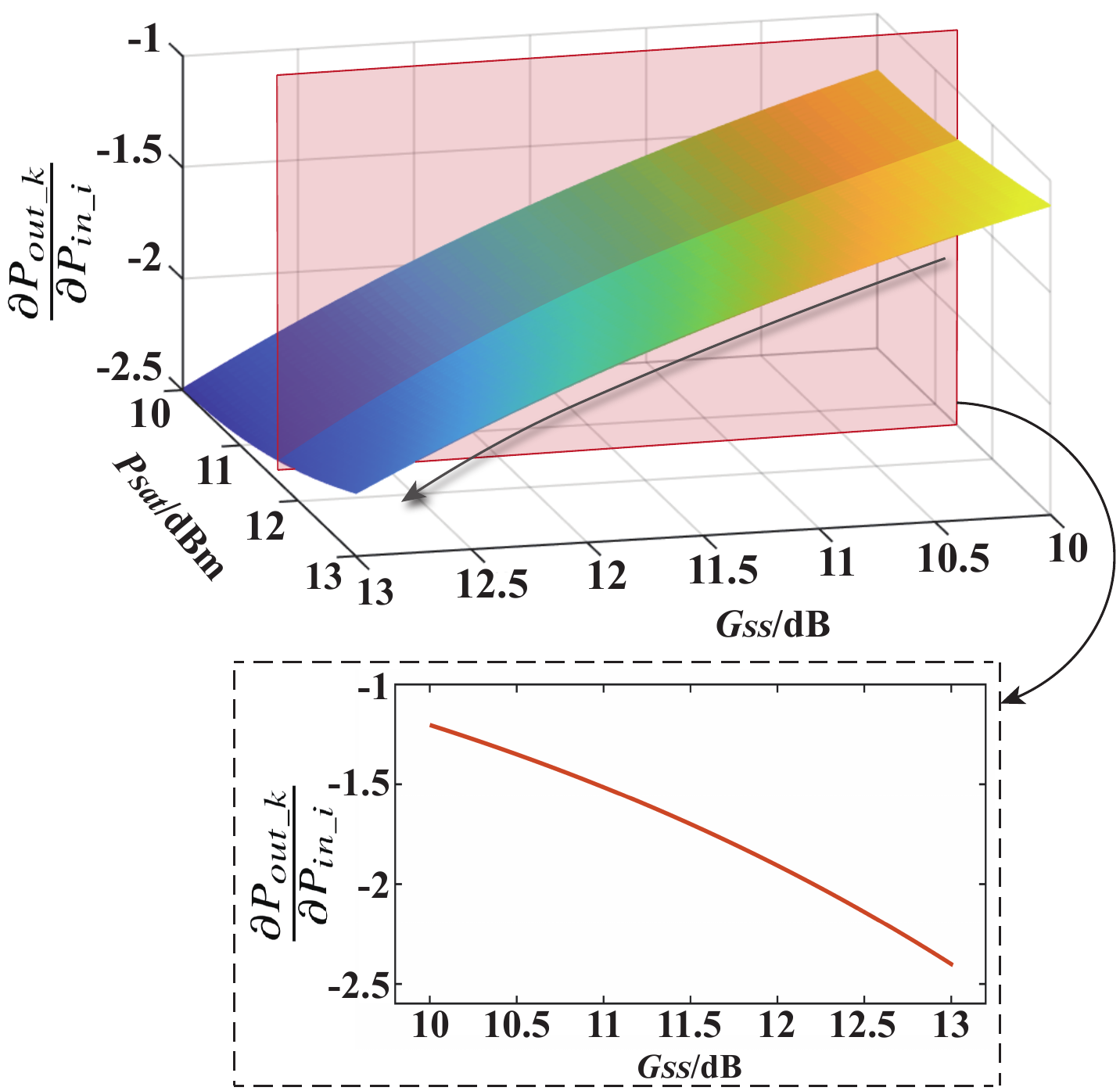}
\caption{The term $\frac{{\partial {P_{out\_k}}}}{{\partial {P_{in\_i}}}}$ evaluates the crosstalk level brought by the ${i^{th}}$ channel. The x-axis and the y-axis are ${G_{ss}}$ and ${P_{sat}}$ respectively, which are the two parameters affect the crosstalk level. The red box indicates the origin of the inset on the right. It is obvious that the interchannel crosstalk level increases with ${G_{ss}}$.}
\label{ct-eva}
\end{figure}

The new BP algorithm aims to alleviate or even annul performance degradation caused by interchannel crosstalk as the device integration level increases. Therefore, the factors influencing the crosstalk level of SOA-based MNS must be investigated. As shown in Eq. ~ (\ref{gain-saturation}) to Eq. (\ref{output-k}), the output of the ${k^{th}}$ channel, ${P_{out\_k}}$, changes with the input of the other channels even if ${P_{in\_k}}$ remains constant. Based on the partial derivative of ${P_{out\_k}}$ to ${P_{in\_i}}$, the crosstalk level of the ${i^{th}}$ channel brought to ${k^{th}}$ can be evaluated exactly at the point where ${P_{out\_k}}$ is affected by ${P_{in\_i}}$. The results of the partial derivative for the gain saturation are shown in Fig. \ref{ct-eva}. As the two parameters ${G_{ss}}$ and ${P_{sat}}$ are set as the x-axis and y-axis and $\frac{{\partial {P_{out\_k}}}}{{\partial {P_{in\_i}}}}$ is on the z-axis, interchannel crosstalk becomes increasingly severe when ${G_{ss}}$ increases. To compare the performance of the proposed ONN under different crosstalk levels, three ${G_{ss}}$ values (${G_{ss}}$ = 20, 23, and 26dB) are used to represent the low, medium, and high crosstalk levels. The value of ${P_{sat}}$ remains unchanged during training.

\begin{figure*}[htbp]
\centering\includegraphics[width=13cm]{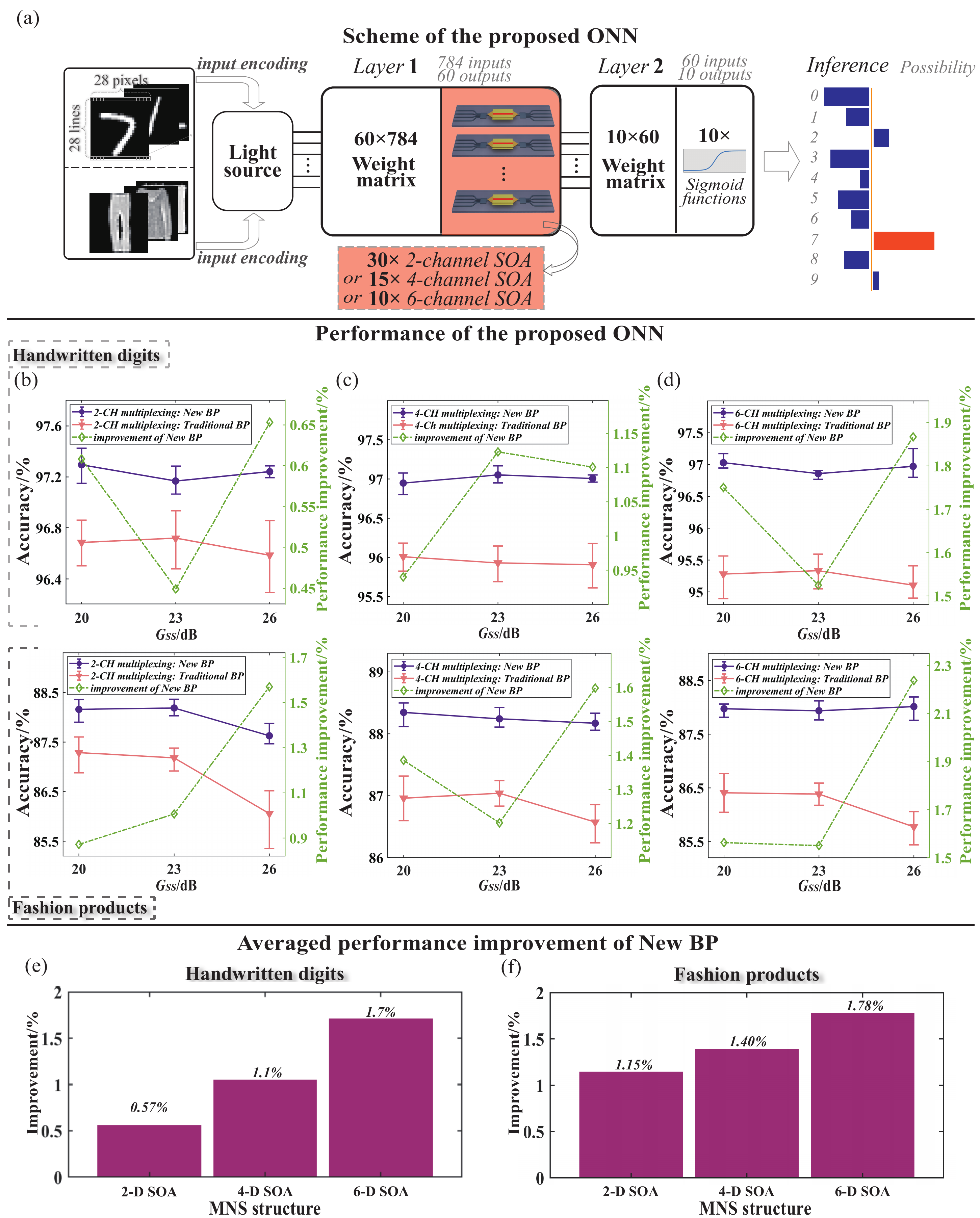}
\caption{(A) The scheme of the proposed ONN for simulation. (B)-(D) The performance of the proposed ONN with 2-channel,4-channel and 6-channel multiplexing SOAs. The x-axis indicates the crosstalk level. The proposed ONN trained by the new BP algorithm demonstrates a steady performance as the crosstalk level and the number of multiplexed channels increases. The one trained by the traditional BP algorithm suffers performance degradation induced by interchannel crosstalk. (E)-(F) The performance improvement of the new BP algorithm over the traditional one rises as more channels of SOAs in the proposed ONN are multiplexed. The new BP algorithm shows significant relevance to larger ONN network with denser-multiplexed MNS structure.}
\label{perf-2d-4d-1}
\end{figure*}

\section{Results}
An ONN architecture involving SOA-based MNS structures was trained using the new algorithm. Simulation results based on the traditional BP algorithm and those under different crosstalk levels were obtained for performance comparison. The applicability of the new BP algorithm was evaluated for an architecture with MNS. 

A schematic of the proposed ONN is presented in Fig. \ref{perf-2d-4d-1}(A). It should be emphasized that it is not realistic to have a photonic circuit with 784 physical inputs and 60 fully connected nodes based on the current cutting-edge practices. However, there are various strategies for dividing the system (or chip) into smaller multiplexed parts to verify the proposed network, as many studies have illustrated. The hidden layer respectively utilizes 2-channel, 4-channel or 6-channel multiplexing SOAs as the MNS. The output layer utilizes a traditional electrically realized sigmoid function, which is common in existing on-chip ONN\cite{shen2017deep}. The corresponding network scale of ONN architecture is set to 784 inputs, 60 neurons in the hidden layer, and 10 neurons in the output layer. Considering the scaling factor of MNS, the number of devices in the hidden layer decreases by half, three-fourths, five-sixths, or even more if more channels of the SOA are multiplexed. 

\subsection{Performance analysis}
Two classification tasks were assigned for the performance analysis based on two datasets: MNIST handwritten digits and fashion-MNIST. In Figs. \ref{perf-2d-4d-1}(B)-(D), the performance of the proposed ONN with 2-channel, 4-channel and 6-channel multiplexing SOAs as MNS is shown. The upper and lower rows of Figs. \ref{perf-2d-4d-1}(B)-(D) correspond to the task of MNIST handwritten digits and fashion-MNIST, respectively. In each figure of Figs. \ref{perf-2d-4d-1}(B)-(D), the solid-line with round marks comes from the result of the proposed ONN that is trained by the new BP training algorithm. For comparison, the solid line with triangular marks represents the result of training using the traditional BP training algorithm. The x-axis indicates the crosstalk level, and the left y-axis indicates the classification accuracy after training.

If the proposed ONN is trained using the new BP algorithm, the individual figures shown in Fig. \ref{perf-2d-4d-1}(B)-(D) shows that the classification accuracy varies slightly under different crosstalk level. In addition, referring to the figures in the row, the performances of the 2-channel to 6-channel multiplexing SOAs are similar. However, as shown by the solid line with triangular marks, blindly improving the integration level through WDM without utilizing the new algorithm decreases the classification accuracy substantially. These trends not only prove the strong resistance of the new BP algorithm to crosstalk but also demonstrate that a denser-multiplexed MNS can be realized without substantial performance degradation with the help of the new BP algorithm. 

For each proposed ONN composed of n-channel(n = 2, 4, or 6) multiplexing SOAs, the training accuracy of the new BP algorithm under different crosstalk levels was summed and averaged, similar to that of the traditional BP algorithm. The gap between these two values, which can be defined as an improvement factor, indicates a performance improvement when the proposed ONN with an n-channel MNS is trained by the new BP algorithm. From another perspective, the necessity for a new BP algorithm for the proposed ONN with an n-channel MNS can be evaluated using this factor. In Figs. \ref{perf-2d-4d-1}(E) and (F), the improvement factor increases with the multiplexed level of SOAs. It is obvious that our new BP algorithm strongly alleviates the problem caused by parallel signal processing in nonlinear devices, and this becomes a necessity when a denser-multiplexed MNS (SOAs with more channels multiplexed in this case) is employed in WDM-ONNs. 

\begin{figure*}[htbp]
\centering\includegraphics[width=15cm]{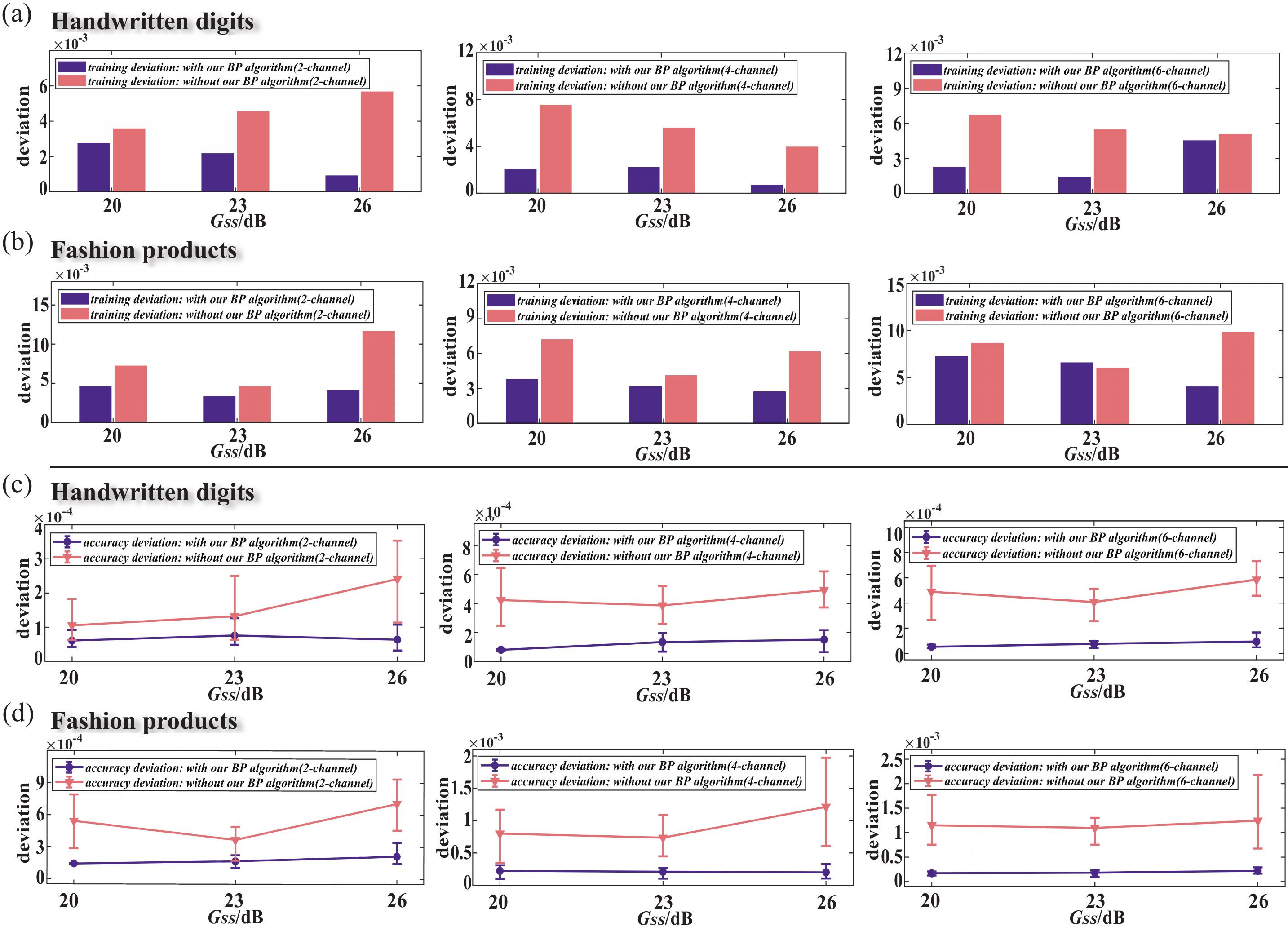}
\caption{(A)-(B) The training deviation of the proposed ONN for both the MNIST handwritten digits and fashion-MNIST classification tasks. The training deviation of the proposed ONN with n-channel(n = 2, 4 or 6) multiplexing SOAs is separately shown in a row. (C)-(D) The accuracy deviation of the proposed ONN with n-channel(n = 2, 4 or 6) multiplexing SOAs separately shown in a row. The x-axis indicates the crosstalk level in (A)-(D). The lower training deviation and accuracy deviation prove that the traditional BP algorithm results in the proposed ONN to converge along the direction deviating from the gradient.}
\label{zd-h}
\end{figure*}

The stability of the new algorithm against interchannel crosstalk arises from the fact that it includes errors induced by crosstalk in the process of BP. In other words, if the crosstalk can be measured (formulized in this case), the algorithm considers it and maintains its performance. The more accurately the crosstalk is measured, the better the performance. However, as indicated by the dashed line with triangular marks, blindly improving the integration level through WDM without utilizing the new algorithm greatly decreases the classification accuracy. The green dashed line, together with the right y-axis, directly indicates the performance improvement caused by the new algorithm.  
The training deviation is defined as the difference between the maximum and minimum accuracies of 10 repetitive training processes of a certain ONN. In Figs. \ref{zd-h}(A) and (B), the training deviation of the proposed ONN trained by the new BP algorithm and traditional BP algorithm for both classification tasks is shown. The results of the n-channel multiplexing SOAs(n = 2, 4, 6) are presented in a row. In most cases, the training deviation of the proposed ONN trained using the new BP algorithm is lower than that of the ONN trained using the traditional BP algorithm. The accuracy deviation, which is defined as the fluctuation in accuracy during an individual training process of a certain ONN, is shown in Figs. \ref{zd-h}(C) and (D) for both classification tasks. If the standard deviation of the accuracy of the last 10 iteration steps during individual training is considered, the accuracy deviation of the proposed ONN trained by the new BP algorithms is shown to be much lower than that trained by the traditional BP algorithm, regardless of the crosstalk levels and the number of multiplexed channels of the SOAs.

These two phenomena indicate that the proposed ONN trained using the traditional BP algorithm does not converge as well as the ONN trained using the new algorithm. As the error induced by crosstalk is not considered in the traditional BP algorithm, the cost function does not descend along the gradient direction. Consequently, the convergence of the network to the global minimum is a random process. In addition, with an increase in the crosstalk level and number of multiplexed channels of SOAs, the descending direction of the cost function further deviates from the gradient direction. Although the randomness caused by the traditional BP algorithm may not result in a markedly larger training or accuracy deviation, as shown In Fig. \ref{zd-h}(B), since we only take finite number of simulations, it is a fatal drawback of the traditional BP algorithm.


\subsection{Power consumption and integration level prospects}
The performance maintenance ability of the proposed ONN and new BP algorithm was proved using the data presented in the previous section. Therefore, it is fair to discuss the advantages of this combination over traditional ONNs. A direct advantage is the elimination of the number of devices used in the nonlinear activation. Both the scaling of integration and flexibility of signal routing are beneficial. However, from the perspective of energy saving, signals are combined in the MNS so that the required input power of each channel in the MNS could be multiple times lower than that of the traditional optical neuron to access the nonlinear operation regime. In other words, light sources can be replaced by low-power sources. For MNS realized by SOA, the power consumption of the nonlinear activation part is also reduced. The basis for the power consumption analysis below is the noncoherent WDM situation for SOA-based MNS, which complies with the architecture presented in the work and the weights are at the power level as usually conducted.

\begin{figure}[htbp]
\centering\includegraphics[width=8.5cm]{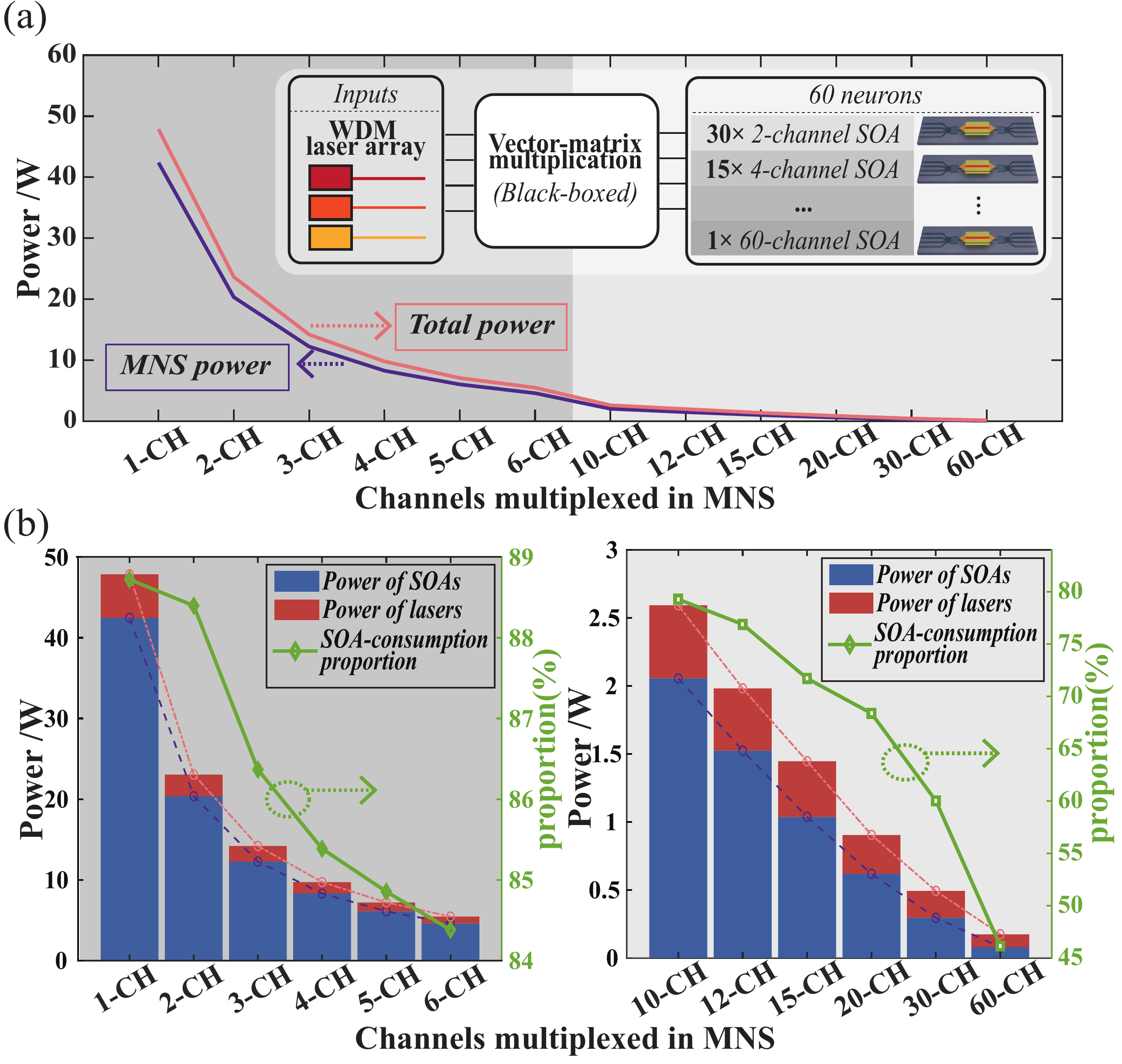}
\caption{ (A) The total power consumption and MNS power consumption of a specific layer with 60 neurons is shown. (B) The detailed proportion of energy consumption is shown. The denser-multiplexed MNS not only lowers the overall power consumption but also occupies less in total power consumption.}
\label{power}
\end{figure}

Based on the aforementioned principles, we theoretically analyzed the power consumption of a specific layer with 60 neurons in the proposed ONN, as shown in the inset of Fig. \ref{power}(A). The consumption induced by vector-matrix multiplication can be seen as a black box with a constant insertion loss factor, which is very common in mainstream ONNs composed of passive devices. The equations Eq. ~ (\ref{gain-saturation}) - Eq. (\ref{output-k}) used in the previous simulation were applied in the analysis, and the external quantum efficiency ${\eta = 0.6}$ was also applied.

\begin{equation}
{P_{SOA\_m}} = \frac{{\sum\limits_{k = 1}^n {{P_{out\_k}}}  - \sum\limits_{k = 1}^n {{P_{in\_k}}} }}{\eta }
\label{Psoa_m}
\end{equation}

\begin{equation}
P = \sum\limits_{m = 1}^M {{P_{out\_m}}}
\label{Psoa}
\end{equation}

\noindent here ${M}$ is defined as the number of SOAs utilized in the MNS structure of this layer. The laser power consumption was  estimated using the external quantum efficiency. 

In Fig. \ref{power}(A), the decrease of total power consumption in WDM-ONNS is obviously shown by a factor of ten as the multiplexed channels of SOAs increase, no matter whether the SOA part is examined individually or together with the light source part. Furthermore, if we separate Fig. \ref{power}(A) into two parts, as shown in Fig. \ref{power}(B), we can clearly read and analyze the proportion that the SOA part occupies in the total power consumption. The green line with square marks indicates that the proportion of the SOA decreases as the multiplexed channels of SOAs increases. In other words, in addition to the total energy-saving property, the proposed ONN with a denser multiplexed MNS structure has great potential for eliminating the proportion of the power consumption of the network's nonlinear activation functions, which is usually realized by active high-power-consumption devices. The general advantage of ONNs over their electrical counterparts was further enhanced by the proposed MNS structure.

\section{Discussion}
We proposed a WDM structure called MNS that can be implemented by various nonlinear devices to improve the parallelism of ONN and a corresponding BP training algorithm to alleviate or even annul the influence of the inevitable interchannel crosstalk caused by the high parallelism of MNS. The performance comparison proves that the combination of the proposed MNS-based WDM-ONN and the new BP algorithm provides markedly similar performance to traditional ONNs, while the footprint of the physical system is decreased. In addition, the power consumption of MNS-based WDM-ONN greatly decreased by a factor of ten as the parallelism of MNS increased. These results proves that our work paves the way for a new type of ONN architecture with smaller scale and lower energy consumption. In addition, our work is demonstrated at a highly abstract level and thus sets up a paradigm for numerous future studies.

\section*{Acknowledgments}
Y. F. Liu thanks to Dr. Ling-Fang Wang and Mr. Chen-Hao Lu for their knowledge of SOA and laser dynamics. Y. F. Liu also thanks Dr. Bei Chen for her instructions and knowledge of ONN setup.

\subsection*{Author Contributions} 
Y. F. Liu and C. Y. Jin jointly conceived the study idea. Y. F. Liu and R. Y. Ren finished the coding of the new BP algorithm. K. J. Huang analyzed the crosstalk induced performance degradation of AI networks. C. Y. Jin, Y. F. Liu, C. H. Li, D. B. Hou, B. W. Wang, and H. Z. Weng analyzed and established the gain-saturation model of SOA. Y. F. Liu wrote the paper with input from the other authors. C. Y. Jin, C. H. Li, F. Liu, and X. Lin supervised the study.

\subsection*{Funding}
This study was supported by the National Key Research and Development Program of China. 
(2021YEB2800500), and the National Natural Science Foundation of China (61574138, 61974131). 
Natural Science Foundation of Zhejiang Province (LGJ21F050001); Major Scientific Project. 
of Zhejiang Laboratory (2019MB0AD01).

\subsection*{Conflicts of Interest}
 The authors declare that there is no conflict of interest regarding the publication of this article.
\subsection*{Data Availability}
The code used in the experiments is open-source and available on GitHub: https://github.com/YifengLiu-ZJU/MNS\_ONN. Classification task datasets are presented in Refs. \cite{xiao2017fashion,deng2012mnist}. Other underlying data are not publicly available at this time but may be obtained from the authors upon reasonable request.

\section*{Supplementary Materials}
\subsection*{The corresponding BP training algorithm}

For the MNS constructed by crosstalk-devices, the output of the i-th channel can be abstracted as multivariable functions in the form of ${y_i} = {f_i}({x_1},{x_2},...,{x_n})$, where ${x_n}$ represents the input of the n-th channel and ${y_i}$ represents the output of the i-th channel. 

\begin{figure}[htbp]
\centering\includegraphics[width=8.5cm]{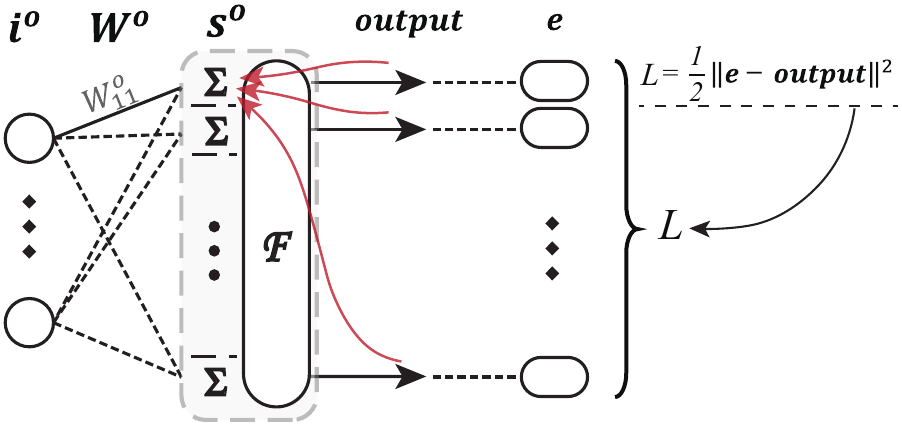}
\caption{A scheme of the output layer in the FCNN. The grey box represents the MNS where the crosstalk is considered. The nonlinear activation function with multiple inputs is $\bm{F}$. During training, each output vector has an expected value, and according to the actual output and expected output, the cost function is derived as  $L$.}
\label{bp-out}
\end{figure}

\begin{figure}[tb]
\centering\includegraphics[width=8cm]{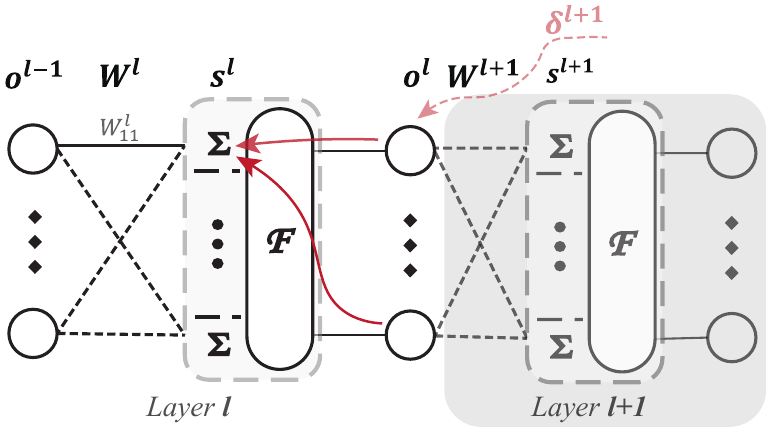}
\caption{A scheme of the hidden layers in the FCNN. The same MNS structure shown in Fig. \ref{bp-out} applies. \emph{Layer l} is the target training layer in this scheme and \emph{Layer l+1} is the output layer if \emph{Layer l} is the rightmost hidden layer. During training, we assume the error of \emph{Layer l+1} is obtained as $\bm{{\delta ^{l+1}}}$ and is propagating to \emph{Layer l} to calculate the error $\bm{{\delta ^l}}$.}
\label{bp-hid}
\end{figure}

According to the position of the layer, we divided the FCNN into two parts: the output layer and hidden layers. Schematic diagrams are shown in Figs. \ref{bp-out} and \ref{bp-hid}, respectively. Both figures involve MNS in the gray box, and the new BP algorithm is illustrated based on them. Although there may be several hidden layers in an FCNN, they all play the same role in receiving the input and passing the output to the next layer after an operation. However, the output layer is the edge of the FCNN and its output is also the output of the FCNN. During the training stage, the error which backpropagates in the FCNN, is generated in the output layer and propagated between the hidden layers. Here, we first illustrate our new BP algorithm in the output layer and then in the hidden layers.

As shown in Fig. \ref{bp-out}, the input of the output layer is $\bm{{i^o}}$, which is also the output of the previous hidden layer. The weight matrix is $\bm{{W^o}}$; thus, the input vector of the activation function can be expressed as $\bm{{s^o}} = \bm{{W^o}} \cdot \bm{{i^o}}$. After the nonlinear activation, the output vector with crosstalk has the following elements: 

\begin{equation}
\bm{output} = \left( \begin{array}{l}
outpu{t_1}\\
outpu{t_2}\\
...\\
outpu{t_n}
\end{array} \right) = \left( \begin{array}{l}
{f_1}(s_1^o,s_2^o,...,s_n^o)\\
{f_2}(s_1^o,s_2^o,...,s_n^o)\\
...\\
{f_n}(s_1^o,s_2^o,...,s_n^o)
\end{array} \right) = \bm{f}(\bm{{s^o}}).
\label{output-of-output}
\end{equation}

\noindent During training, the expected outputs is given as $\bm{{e}}$, and the cost function can be defined as

\begin{equation}
L = \frac{1}{2}{\left\| {\bm{e} - \bm{output}} \right\|^2}.
\label{cost-function}
\end{equation}

\noindent As the weight matrix $\bm{{W^o}}$ updates, the gradient descent method is applied

\begin{equation}
\bm{W^o} = \bm{W^o} + \eta \times \frac{{\partial L}}{{\partial \bm{W^o}}}.
\label{update-w}
\end{equation}

\noindent Here $\eta$ represents the learning rate.

According to the chain rule, we derive the following formula for the gradients:

\begin{equation}
\frac{{\partial L}}{{\partial \bm{{W^o}}}} = \frac{{\partial L}}{{\partial \bm{{s^o}}}} \cdot \frac{{\partial \bm{{s^o}}}}{{\partial \bm{{W^o}}}} = \frac{{\partial L}}{{\partial \bm{{s^o}}}} \cdot \frac{{\partial \bm{{W^o}} \cdot \bm{{i^o}}}}{{\partial \bm{{W^o}}}} = \frac{{\partial L}}{{\partial \bm{{s^o}}}} \cdot (\bm{{i^o}})^T
\label{chain-1}
\end{equation}

\begin{equation}
\frac{{\partial L}}{{\partial \bm{{W^o}}}} = {(\begin{array}{*{20}{c}}
{\frac{{\partial L}}{{\partial s_1^o}}}&{\frac{{\partial L}}{{\partial s_2^o}}}&{...}&{\frac{{\partial L}}{{\partial s_n^o}}}
\end{array})^T} \cdot (\bm{{i^o}})^T \buildrel \Delta \over = \bm{{\delta ^o}} \cdot (\bm{{i^o}})^T
\label{chain-2}
\end{equation}

\noindent where the vector $\bm{{\delta ^o}}$ is called the error. If we consider the individual elements of $\bm{{\delta ^o}}$ such as $\frac{{\partial L}}{{\partial s_1^o}}$, the special part of the new BP algorithm that addresses crosstalk is already involved in the two equations above. The element-wise expansion of Eq. ~ (\ref{cost-function}) is 

\begin{equation}
L = \frac{1}{2}\left[ {{{\left( {{e_1} - outpu{t_1}} \right)}^2} + ... + {{\left( {{e_n} - outpu{t_n}} \right)}^2}} \right].
\label{cost-function-ex}
\end{equation}

\noindent According to Eq. (\ref{output-of-output}), and Eq. (\ref{cost-function}), partial $\frac{{\partial L}}{{\partial s_1^o}}$ is expressed as follows:

\begin{equation}
\frac{{\partial L}}{{\partial s_1^o}} = \left[ {(outpu{t_1} - {e_1}) \cdot \frac{{\partial outpu{t_1}}}{{\partial s_1^o}} + ... + (outpu{t_n} - {e_n}) \cdot \frac{{\partial outpu{t_n}}}{{\partial s_1^o}}} \right].
\label{part-s1o-ex}
\end{equation}

In contrast to the traditional BP algorithm, the proposed BP algorithm requires the calculation of partial derivatives from $\frac{{\partial outpu{t_1}}}{{\partial s_1^o}}$ to $\frac{{\partial outpu{t_n}}}{{\partial s_1^o}}$ so that the cost can descend along the gradient correctly. The red arrows in Fig. \ref{bp-out} visualize the BP of the partial derivatives according to the chain rule. In vector form, we can write Eq. ~ (\ref{part-s1o-ex}) as 

\begin{equation}
\frac{{\partial L}}{{\partial s_1^o}} = {\left( {\frac{{\partial \bm{output}}}{{\partial s_1^o}}} \right)^T} \cdot (\bm{output} - \bm{e}).
\label{part-s1o-sl}
\end{equation}

\noindent Because the first element of $\bm{{\delta ^o}}$ is derived in Eq. (\ref{part-s1o-sl}), the other elements can be similarly derived. We directly provide an expression of $\bm{{\delta ^o}}$ composed of all the other elements in the following form:
\begin{equation}
\bm{{\delta ^o}} = \frac{{\partial L}}{{\partial \bm{{s^o}}}} = {\left( {\begin{array}{*{20}{c}}
{{{\left( {\frac{{\partial \bm{output}}}{{\partial s_1^o}}} \right)}^T}}&{...}&{{{\left( {\frac{{\partial \bm{output}}}{{\partial s_n^o}}} \right)}^T}}
\end{array}} \right)^T} \cdot (\bm{output} - \bm{e}) = \left( {\frac{{\partial \bm{output}}}{{\partial \bm{{s^o}}}}} \right) \cdot (\bm{output} - \bm{e}).
\label{part-s1}
\end{equation}

At the right end of Eq. (\ref{part-s1}), the expression of $\bm{{\delta ^o}}$ is tightly coupled with the expression of the traditional BP algorithm. However, we must bear in mind the inner differences caused by crosstalk between our new BP algorithm and the traditional algorithm. To further emphasize this difference, we provide an element-wise expanded expression of ${\left( {\frac{{\partial \bm{output}}}{{\partial \bm{{s^o}}}}} \right)}$ in matrix form, as used in the main text. 

\begin{equation}
{\frac{{\partial \bm{output}}}{{\partial \bm{{s^o}}}}}  = \left( {\begin{array}{*{20}{c}}
{\frac{{\partial outpu{t_1}}}{{\partial s_1^o}}}& \ldots &{\frac{{\partial outpu{t_n}}}{{\partial s_1^o}}}\\
 \vdots & \ddots & \vdots \\
{\frac{{\partial outpu{t_1}}}{{\partial s_n^o}}}& \cdots &{\frac{{\partial outpu{t_n}}}{{\partial s_n^o}}}
\end{array}} \right).
\label{ele-expand_1}
\end{equation}

Each element in the matrix has a definition corresponding to the crosstalk among channels. In a traditional BP algorithm, the elements on the diagonal have definitions, whereas the off-diagonal elements are left undefined. The new BP algorithm deals with physical crosstalk and couples mathematical operations to the undefined items of the traditional BP algorithm.

For the hidden layer shown in Fig. (\ref{bp-hid}), the error is backpropagated from \emph{ layer l+1} to \emph{ layer l}. Here, $\bm{{o^{l-1}}}$ is the input to \emph{Layer l} and output to \emph{Layer l-1}. The weight matrix $\bm{{W^l}}$ is then multiplexed to obtain the output of the linear part of \emph{Layer l}, $\bm{{s^{l}}} = \bm{{W^l}} \cdot \bm{{o^{l - 1}}}$. Nonlinear activation is performed in the MNS in the gray-dashed box, and the output of \emph{Layer l} is obtained as $\bm{{o^l}} = \bm{f}(\bm{{s^l}})$ according to Eq. (\ref{output-of-output}). Now, if the weight matrix $\bm{{W^l}}$ is updated according to Eq. (\ref{update-w}), $\frac{{\partial L}}{{\partial \bm{{W^l}}}}$ must be obtained using the following approach: 

\begin{equation}
\frac{{\partial L}}{{\partial \bm{{W^l}}}} = \frac{{\partial L}}{{\partial \bm{{s^{{\rm{l}}}}}}} \cdot \frac{{\partial \bm{{s^l}}}}{{\partial \bm{{W^l}}}} = \frac{{\partial \bm{f}(\bm{{s^l}})}}{{\partial \bm{{s^l}}}} \cdot \frac{{\partial \bm{{s^{{\rm{l + 1}}}}}}}{{\partial \bm{f}(\bm{{s^l}})}} \cdot \frac{{\partial L}}{{\partial \bm{{s^{{\rm{l + 1}}}}}}} \cdot \frac{{\partial \bm{{s^l}}}}{{\partial \bm{{W^l}}}}.
\label{chain-3}
\end{equation}

\noindent Combined with Eq. (\ref{chain-1}) and Eq. (\ref{chain-2}), 
\begin{equation}
\frac{{\partial L}}{{\partial \bm{{W^l}}}} = \frac{{\partial \bm{f}(\bm{{s^l}})}}{{\partial \bm{{s^l}}}} \cdot (\bm{{W^{l + 1}}})^T \cdot \bm{{\delta ^{l + 1}}} \cdot {(\bm{{o^{l - 1}}})^T}.
\label{chain-4}
\end{equation}

\noindent Here $\bm{{\delta ^{l + 1}}}$ is the error of \emph{Layer l+1} backpropagating through the weight matrix to \emph{Layer l}. The term $\frac{{\partial \bm{f}(\bm{{s^l}})}}{{\partial \bm{{s^l}}}}$ is a matrix with elements defined by the interchannel crosstalk 

\begin{equation}
\frac{{\partial \bm{f}(\bm{{s^l}})}}{{\partial \bm{{s^l}}}} = \left( {\begin{array}{*{20}{c}}
{\frac{{\partial {{\rm{f}}_1}({s^l})}}{{\partial s_1^l}}}& \ldots &{\frac{{\partial {{\rm{f}}_n}({s^l})}}{{\partial s_1^l}}}\\
 \vdots & \ddots & \vdots \\
{\frac{{\partial {{\rm{f}}_1}({s^l})}}{{\partial s_n^l}}}& \cdots &{\frac{{\partial {{\rm{f}}_n}({s^l})}}{{\partial s_n^l}}}.
\end{array}} \right)
\label{ele-expand-2}
\end{equation}

\noindent To date, the whole process of the new BP algorithm has been elaborated. The weight matrix in each layer is updated according to the process described above.

\printbibliography

@article{xiao2017fashion,
  title={Fashion-mnist: a novel image dataset for benchmarking machine learning algorithms},
  author={Xiao, Han and Rasul, Kashif and Vollgraf, Roland},
  journal={arXiv preprint arXiv:1708.07747},
  year={2017}
}

@article{mnih2015human,
  title={Human-level control through deep reinforcement learning},
  author={Mnih, Volodymyr and Kavukcuoglu, Koray and Silver, David and Rusu, Andrei A and Veness, Joel and Bellemare, Marc G and Graves, Alex and Riedmiller, Martin and Fidjeland, Andreas K and Ostrovski, Georg and others},
  journal={nature},
  volume={518},
  number={7540},
  pages={529--533},
  year={2015},
  publisher={Nature Publishing Group}
}

@article{silver2017mastering,
  title={Mastering the game of go without human knowledge},
  author={Silver, David and Schrittwieser, Julian and Simonyan, Karen and Antonoglou, Ioannis and Huang, Aja and Guez, Arthur and Hubert, Thomas and Baker, Lucas and Lai, Matthew and Bolton, Adrian and others},
  journal={nature},
  volume={550},
  number={7676},
  pages={354--359},
  year={2017},
  publisher={Nature Publishing Group}
}

@article{butler2018machine,
  title={Machine learning for molecular and materials science},
  author={Butler, Keith T and Davies, Daniel W and Cartwright, Hugh and Isayev, Olexandr and Walsh, Aron},
  journal={Nature},
  volume={559},
  number={7715},
  pages={547--555},
  year={2018},
  publisher={Nature Publishing Group}
}

@article{de2019photonic,
  title={Photonic neural networks: A survey},
  author={De Marinis, Lorenzo and Cococcioni, Marco and Castoldi, Piero and Andriolli, Nicola},
  journal={IEEE Access},
  volume={7},
  pages={175827--175841},
  year={2019},
  publisher={IEEE}
}

@article{xu2021survey,
  title={A survey of approaches for implementing optical neural networks},
  author={Xu, Runqin and Lv, Pin and Xu, Fanjiang and Shi, Yishi},
  journal={Optics \& Laser Technology},
  volume={136},
  pages={106787},
  year={2021},
  publisher={Elsevier}
}

@article{shen2017deep,
  title={Deep learning with coherent nanophotonic circuits},
  author={Shen, Yichen and Harris, Nicholas C and Skirlo, Scott and Prabhu, Mihika and Baehr-Jones, Tom and Hochberg, Michael and Sun, Xin and Zhao, Shijie and Larochelle, Hugo and Englund, Dirk and others},
  journal={Nature photonics},
  volume={11},
  number={7},
  pages={441--446},
  year={2017},
  publisher={Nature Publishing Group}
}

@article{farhat1985optical,
  title={Optical implementation of the Hopfield model},
  author={Farhat, Nabil H and Psaltis, Demetri and Prata, Aluizio and Paek, Eung},
  journal={Applied optics},
  volume={24},
  number={10},
  pages={1469--1475},
  year={1985},
  publisher={Optica Publishing Group}
}

@article{gruber2000planar,
  title={Planar-integrated optical vector-matrix multiplier},
  author={Gruber, Matthias and Jahns, J{\"u}rgen and Sinzinger, Stefan},
  journal={Applied optics},
  volume={39},
  number={29},
  pages={5367--5373},
  year={2000},
  publisher={Optica Publishing Group}
  }

@article{feldmann2019all,
  title={All-optical spiking neurosynaptic networks with self-learning capabilities},
  author={Feldmann, Johannes and Youngblood, Nathan and Wright, C David and Bhaskaran, Harish and Pernice, Wolfram HP},
  journal={Nature},
  volume={569},
  number={7755},
  pages={208--214},
  year={2019},
  publisher={Nature Publishing Group}
}

@article{lin2018all,
  title={All-optical machine learning using diffractive deep neural networks},
  author={Lin, Xing and Rivenson, Yair and Yardimci, Nezih T and Veli, Muhammed and Luo, Yi and Jarrahi, Mona and Ozcan, Aydogan},
  journal={Science},
  volume={361},
  number={6406},
  pages={1004--1008},
  year={2018},
  publisher={American Association for the Advancement of Science}
}

@inproceedings{ishihara2019optical,
  title={An optical neural network architecture based on highly parallelized WDM-multiplier-accumulator},
  author={Ishihara, Tohru and Shiomi, Jun and Hattori, Naoki and Masuda, Yutaka and Shinya, Akihiko and Notomi, Masaya},
  booktitle={2019 IEEE/ACM Workshop on Photonics-Optics Technology Oriented Networking, Information and Computing Systems (PHOTONICS)},
  pages={15--21},
  year={2019},
  organization={IEEE}
}

@article{totovic2022programmable,
  title={Programmable photonic neural networks combining WDM with coherent linear optics},
  author={Totovic, Angelina and Giamougiannis, George and Tsakyridis, Apostolos and Lazovsky, David and Pleros, Nikos},
  journal={Scientific reports},
  volume={12},
  number={1},
  pages={1--13},
  year={2022},
  publisher={Nature Publishing Group}
}

@article{mourgias2020all,
  title={All-optical WDM recurrent neural networks with gating},
  author={Mourgias-Alexandris, George and Dabos, George and Passalis, Nikolaos and Totovi{\'c}, Angelina and Tefas, Anastasios and Pleros, Nikos},
  journal={IEEE Journal of Selected Topics in Quantum Electronics},
  volume={26},
  number={5},
  pages={1--7},
  year={2020},
  publisher={IEEE}
}

@article{shi2019deep,
  title={Deep neural network through an InP SOA-based photonic integrated cross-connect},
  author={Shi, Bin and Calabretta, Nicola and Stabile, Ripalta},
  journal={IEEE Journal of Selected Topics in Quantum Electronics},
  volume={26},
  number={1},
  pages={1--11},
  year={2019},
  publisher={IEEE}
}

@article{feldmann2021parallel,
  title={Parallel convolutional processing using an integrated photonic tensor core},
  author={Feldmann, Johannes and Youngblood, Nathan and Karpov, Maxim and Gehring, Helge and Li, Xuan and Stappers, Maik and Le Gallo, Manuel and Fu, Xin and Lukashchuk, Anton and Raja, Arslan Sajid and others},
  journal={Nature},
  volume={589},
  number={7840},
  pages={52--58},
  year={2021},
  publisher={Nature Publishing Group}
}

@article{xu202111,
  title={11 TOPS photonic convolutional accelerator for optical neural networks},
  author={Xu, Xingyuan and Tan, Mengxi and Corcoran, Bill and Wu, Jiayang and Boes, Andreas and Nguyen, Thach G and Chu, Sai T and Little, Brent E and Hicks, Damien G and Morandotti, Roberto and others},
  journal={Nature},
  volume={589},
  number={7840},
  pages={44--51},
  year={2021},
  publisher={Nature Publishing Group}
}

@article{xu2020photonic,
  title={Photonic perceptron based on a Kerr Microcomb for high-speed, scalable, optical neural networks},
  author={Xu, Xingyuan and Tan, Mengxi and Corcoran, Bill and Wu, Jiayang and Nguyen, Thach G and Boes, Andreas and Chu, Sai T and Little, Brent E and Morandotti, Roberto and Mitchell, Arnan and others},
  journal={Laser \& Photonics Reviews},
  volume={14},
  number={10},
  pages={2000070},
  year={2020},
  publisher={Wiley Online Library}
}

@inproceedings{zhao2019hardware,
  title={Hardware-software co-design of slimmed optical neural networks},
  author={Zhao, Zheng and Liu, Derong and Li, Meng and Ying, Zhoufeng and Zhang, Lu and Xu, Biying and Yu, Bei and Chen, Ray T and Pan, David Z},
  booktitle={Proceedings of the 24th Asia and South Pacific Design Automation Conference},
  pages={705--710},
  year={2019}
}

@article{mourgias2019all,
  title={An all-optical neuron with sigmoid activation function},
  author={Mourgias-Alexandris, George and Tsakyridis, A and Passalis, N and Tefas, Anastasios and Vyrsokinos, K and Pleros, Nikolaos},
  journal={Optics express},
  volume={27},
  number={7},
  pages={9620--9630},
  year={2019},
  publisher={Optical Society of America}
}

@article{shi2022inp,
  title={InP photonic integrated multi-layer neural networks: Architecture and performance analysis},
  author={Shi, Bin and Calabretta, Nicola and Stabile, Ripalta},
  journal={APL Photonics},
  volume={7},
  number={1},
  pages={010801},
  year={2022},
  publisher={AIP Publishing LLC}
}

@article{shastri2021photonics,
  title={Photonics for artificial intelligence and neuromorphic computing},
  author={Shastri, Bhavin J and Tait, Alexander N and Ferreira de Lima, Thomas and Pernice, Wolfram HP and Bhaskaran, Harish and Wright, C David and Prucnal, Paul R},
  journal={Nature Photonics},
  volume={15},
  number={2},
  pages={102--114},
  year={2021},
  publisher={Nature Publishing Group UK London}
}

@article{huang2022prospects,
  title={Prospects and applications of photonic neural networks},
  author={Huang, Chaoran and Sorger, Volker J and Miscuglio, Mario and Al-Qadasi, Mohammed and Mukherjee, Avilash and Lampe, Lutz and Nichols, Mitchell and Tait, Alexander N and Ferreira de Lima, Thomas and Marquez, Bicky A and others},
  journal={Advances in Physics: X},
  volume={7},
  number={1},
  pages={1981155},
  year={2022},
  publisher={Taylor \& Francis}
}

@article{zhou2022photonic,
  title={Photonic matrix multiplication lights up photonic accelerator and beyond},
  author={Zhou, Hailong and Dong, Jianji and Cheng, Junwei and Dong, Wenchan and Huang, Chaoran and Shen, Yichen and Zhang, Qiming and Gu, Min and Qian, Chao and Chen, Hongsheng and others},
  journal={Light: Science \& Applications},
  volume={11},
  number={1},
  pages={30},
  year={2022},
  publisher={Nature Publishing Group UK London}
}

@article{zhang2021optical,
  title={An optical neural chip for implementing complex-valued neural network},
  author={Zhang, Hui and Gu, Mile and Jiang, XD and Thompson, Jayne and Cai, Hong and Paesani, S and Santagati, R and Laing, A and Zhang, Y and Yung, MH and others},
  journal={Nature communications},
  volume={12},
  number={1},
  pages={457},
  year={2021},
  publisher={Nature Publishing Group UK London}
}

@article{qian2020performing,
  title={Performing optical logic operations by a diffractive neural network},
  author={Qian, Chao and Lin, Xiao and Lin, Xiaobin and Xu, Jian and Sun, Yang and Li, Erping and Zhang, Baile and Chen, Hongsheng},
  journal={Light: Science \& Applications},
  volume={9},
  number={1},
  pages={59},
  year={2020},
  publisher={Nature Publishing Group UK London}
}

@article{wang2022optical,
  title={An optical neural network using less than 1 photon per multiplication},
  author={Wang, Tianyu and Ma, Shi-Yuan and Wright, Logan G and Onodera, Tatsuhiro and Richard, Brian C and McMahon, Peter L},
  journal={Nature Communications},
  volume={13},
  number={1},
  pages={123},
  year={2022},
  publisher={Nature Publishing Group UK London}
}

@article{huang2021silicon,
  title={A silicon photonic--electronic neural network for fibre nonlinearity compensation},
  author={Huang, Chaoran and Fujisawa, Shinsuke and de Lima, Thomas Ferreira and Tait, Alexander N and Blow, Eric C and Tian, Yue and Bilodeau, Simon and Jha, Aashu and Yaman, Fatih and Peng, Hsuan-Tung and others},
  journal={Nature Electronics},
  volume={4},
  number={11},
  pages={837--844},
  year={2021},
  publisher={Nature Publishing Group UK London}
}

@article{tait2016microring,
  title={Microring weight banks},
  author={Tait, Alexander N and Wu, Allie X and De Lima, Thomas Ferreira and Zhou, Ellen and Shastri, Bhavin J and Nahmias, Mitchell A and Prucnal, Paul R},
  journal={IEEE Journal of Selected Topics in Quantum Electronics},
  volume={22},
  number={6},
  pages={312--325},
  year={2016},
  publisher={IEEE}
}

@article{wang2022metasurface,
  title={Metasurface on integrated photonic platform: from mode converters to machine learning},
  author={Wang, Zi and Xiao, Yahui and Liao, Kun and Li, Tiantian and Song, Hao and Chen, Haoshuo and Uddin, SM Zia and Mao, Dun and Wang, Feifan and Zhou, Zhiping and others},
  journal={Nanophotonics},
  volume={11},
  number={16},
  pages={3531--3546},
  year={2022},
  publisher={De Gruyter}
}

@article{zhu2022space,
  title={Space-efficient optical computing with an integrated chip diffractive neural network},
  author={Zhu, HH and Zou, Jun and Zhang, Hengyi and Shi, YZ and Luo, SB and Wang, N and Cai, H and Wan, LX and Wang, Bo and Jiang, XD and others},
  journal={Nature communications},
  volume={13},
  number={1},
  pages={1044},
  year={2022},
  publisher={Nature Publishing Group UK London}
}

@article{tait2019silicon,
  title={Silicon photonic modulator neuron},
  author={Tait, Alexander N and De Lima, Thomas Ferreira and Nahmias, Mitchell A and Miller, Heidi B and Peng, Hsuan-Tung and Shastri, Bhavin J and Prucnal, Paul R},
  journal={Physical Review Applied},
  volume={11},
  number={6},
  pages={064043},
  year={2019},
  publisher={APS}
}

@article{williamson2019reprogrammable,
  title={Reprogrammable electro-optic nonlinear activation functions for optical neural networks},
  author={Williamson, Ian AD and Hughes, Tyler W and Minkov, Momchil and Bartlett, Ben and Pai, Sunil and Fan, Shanhui},
  journal={IEEE Journal of Selected Topics in Quantum Electronics},
  volume={26},
  number={1},
  pages={1--12},
  year={2019},
  publisher={IEEE}
}

@article{amin2019ito,
  title={ITO-based electro-absorption modulator for photonic neural activation function},
  author={Amin, Rubab and George, JK and Sun, Shuai and Ferreira de Lima, Thomas and Tait, Alexander N and Khurgin, JB and Miscuglio, Mario and Shastri, Bhavin J and Prucnal, Paul R and El-Ghazawi, Tarek and others},
  journal={APL Materials},
  volume={7},
  number={8},
  pages={081112},
  year={2019},
  publisher={AIP Publishing LLC}
}

@article{liu2021research,
  title={Research progress in optical neural networks: theory, applications and developments},
  author={Liu, Jia and Wu, Qiuhao and Sui, Xiubao and Chen, Qian and Gu, Guohua and Wang, Liping and Li, Shengcai},
  journal={PhotoniX},
  volume={2},
  number={1},
  pages={1--39},
  year={2021},
  publisher={SpringerOpen}
}

@article{sui2020review,
  title={A review of optical neural networks},
  author={Sui, Xiubao and Wu, Qiuhao and Liu, Jia and Chen, Qian and Gu, Guohua},
  journal={IEEE Access},
  volume={8},
  pages={70773--70783},
  year={2020},
  publisher={IEEE}
}

@article{deng2012mnist,
  title={The mnist database of handwritten digit images for machine learning research [best of the web]},
  author={Deng, Li},
  journal={IEEE signal processing magazine},
  volume={29},
  number={6},
  pages={141--142},
  year={2012},
  publisher={IEEE}
}

@article{refadv,
author = {Tiankuang Zhou  and Wei Wu  and Jinzhi Zhang  and Shaoliang Yu  and Lu Fang},
title = {Ultrafast dynamic machine vision with spatiotemporal photonic computing},
journal = {Science Advances},
volume = {9},
number = {23},
pages = {eadg4391},
year = {2023},
doi = {10.1126/sciadv.adg4391},
URL = {https://www.science.org/doi/abs/10.1126/sciadv.adg4391},
eprint = {https://www.science.org/doi/pdf/10.1126/sciadv.adg4391}}

\end{document}